# Repumping ground-state population in a coherently driven atomic resonance

**Asif Sinay*, Moshe Shuker, Ofer Firstenberg, Amnon Fisher, Amit Ben-Kish, Jeff Steinhauer**

*Department of Physics, Technion-Israel institute of technology, Israel*
**asif.sinay@gmail.com*

**Abstract:** We experimentally demonstrate an optical pumping technique to pump a dilute rubidium vapor into the $m_F = 0$ ground states. The technique utilizes selection rules that forbid the excitation of the $m_F = 0$ state by linearly-polarized light. A substantial increase in the transparency contrast of coherent population trapping in the clock transition is demonstrated.

## References and links

1. J. Vanier, "Atomic clocks based on coherent population trapping: a review", Appl. Phys. B 81, 421-442 (2005).
2. E. Arimondo, "Coherent population trapping in laser spectroscopy," Progress in Optics, 35, 257, 1996.
3. J. Vanier, M. W. Levine, D. Janssen, and M. Delaney, "Contrast and linewidth of the coherent population trapping transmission hyperfine resonance line in 87Rb: Effect of optical pumping", PRA 67, 065802 (2003).
4. Y.Y. Jau, A.B. Post, N.N. Kuzma, A.M. Braun, M.V. Romalis, W. Happer, in Proceedings of the 2003 IEEE International Frequency Control Symposium & PDA Exhibition Jointly with the 17th European Frequency and Time Forum, 33 (2003).
5. Y.Y. Jau, A.B. Post, N.N.Kuzma, A.M.Braun,M.V. Romalis, W.Happer, Phys. Rev. Lett. 92, 110 801 (2004).
6. Y.-Y. Jau, E. Miron, A. B. Post, N. N. Kuzma, and W. Happer, "Push-Pull Optical Pumping of Pure Superposition State", Phys. Rev. Lett., 93, 160 802 (2004).
7. T. Zanon, S. Tremine, S. Guerandel, E. De Clercq, D. Holleville, N. Dimarcq, A. Clairon, IEEE Trans. Instrum. Meas. 54, 776 (2005).
8. M. Rosenbluh, V. Shah, S. Knappe, and J. Kitching, "Differentially detected coherent population trapping resonances excited by orthogonally polarized laser fields," Opt. Express 14, 6588-6594 (2006).
9. A.V. Taichenachev, V.I. Yudin, V.L. Velichansky, S.V. Kargapoltsev, R. Wynands, J. Kitching, L. Hollberg, JETP Lett. 80, 23 (2004).
10. S. Kargapoltsev, J. Kitching, L. Hollberg, A. V. Taichenachev, V. L. Velichanski and V. I. Yudin, "High-contrast dark resonance in σ+–σ- optical field," Laser Phys. Lett., 1, 495, 2004.
11. M. Zhu, "High Contrast Signal in a Coherent Population Trapping Based Atomic Frequency Standard Application", M. Zhu, in Proceedings of the 2003 IEEE International Frequency Control Symposium & PDA Exhibition Jointly with the 17th European Frequency and Time Forum, 16 (2003).
12. A. V. Taichenachev, V. I. Yudin, V. L. Velichansky, and S. A. Zibrov , "On the Unique Possibility of Significantly Increasing the Contrastof Dark Resonances on the D1 Line of 87Rb" JETP Letters, Vol. 82, No. 7, 2005, pp. 398–403.
13. Eugeniy E. Mikhailov, Travis Horrom, Nathan Belcher, Irina Novikova, "Performance of a prototype atomic clock based on lin||lin coherent population trapping resonances in Rb atomic vapor", arXiv:0910.5881, (2009).
14. G. Kazakov, I. Mazets, Yu. Rozhdestvensky, G. Mileti, J. Delporte and B. Matisov, "High-contrast dark resonance on D2-line of 87Rb in a vapor cell with different directions of the pump-probe waves", Eur. Phys. J. D 35, 445-448(2005).
15. Daniel A. Steck, Rubidium 87 D Line Data, Theoretical Division (T-8), MS B285 Los Alamos National Laboratory Los Alamos, NM 87545 25 September 2001 (revision 1.6, 14 October 2003).
16. W. Happer, "Optical Pumping", Reviews of Modern Physics, Vol. 44, No. 2. (April 1972), 169.
17. C. Cohen-Tannoudji, J. Dupont-Roc, and G. Grynberg, "Atom-Photon Interactions: Basic Processes and Applications", 1998.
18. S. Knappe, J. Kitching, L. Hollberg, and R. Wynands, "Temperature dependence of coherent population trapping resonances", Appl. Phys. B 74, 217 (2002).
19. I. Novikova, Y. Xiao, D. F. Phillips, and R. L. Walsworth, J. Mod. Opt. 52, 2381 (2005).

## 1. Introduction

A well-studied technique to implement an all-optical atomic frequency standard utilizes a coherent population trapping (CPT) resonance [1,2] within the clock transition, *i.e.* between two, magnetically insensitive, $m_F = 0$ ground-state levels of an alkali atom. The short-term frequency stability is determined by the width and the amplitude of the transparency resonance and by the accompanied noise. The amplitude of the CPT resonance is proportional to the atomic population within the clock transition, and hence it is desirable to increase the population fraction in the relevant $m_F = 0$ states. At thermal equilibrium, the atoms are equally distributed among all possible ground-state sub-levels, and only a small fraction participates in the clock transition. Moreover, the common approach of using a circularly polarized light optically pumps most of the atoms into the highly polarized spin states [3], further reducing the population in the $m_F = 0$ states. While this pumping is useful to end-state schemes, which utilize the absence of spin-exchange relaxation in highly polarized states [4,5], it is harmful to the $m_F = 0$ CPT resonance.

Several techniques have been proposed to increase the CPT contrast by reducing the effect of pumping to highly polarized states. Most of these introduce an additional laser field, with the orthogonal circular polarization, that effectively pumps the atoms back to the central $m_F = 0$ levels, while maintaining most of the CPT coherence. The destructive interference between the induced coherence of the two circular polarizations can be circumvented by temporarily alternating the left and right polarizations [6]; by inducing an appropriate phase delay (lin per lin), either with co-propagating beams [7,8] or with a counter-propagating configuration [9,10]; or utilizing the AC stark shift [11]. An alternative approach utilizes the four $m_F = \pm 1$ Zeeman states, instead of the $m_F = 0$, using a symmetric configuration of left and right circular polarization (lin par lin) [12, 13].

Here we experimentally demonstrate a technique to improve the CPT contrast, by pumping the atoms to the $m_F = 0$ levels with two π-polarized beams (Figs. 1a and 1b). The technique, first introduced in Ref. [14], exploits the forbidden optical transitions ($F = 1$, $m_F = 0$) → ($F' = 1$, $m_F = 0$) and ($F = 2$, $m_F = 0$) → ($F' = 2$, $m_F = 0$), which have a vanishing dipole matrix elements [15]. Using two π-polarized pumping beams, all ground levels except the two $m_F = 0$ are repumped, leading to increased population in the clock transition (depopulation optical pumping [16]). We further investigate additional effects of the repump-beams, such as broadening and off-resonance pumping. In section 2, we describe a theoretical model for CPT in a multi-level atom, incorporating the four optical modes and all the relevant atomic levels. In addition to the CPT process, the model describes population pumping to the end state, ground-state depolarization, and re-pumping of population to the $m_F = 0$ levels. In section 3, we describe the experimental setup and results. We show the predicted increase in contrast, due to the repump-beams, along with a broadening of the CPT line. Finally, we conclude our work in section 4 and discuss the results.

## II. Model system

The full 16-level structure of the D1 transition of $^{87}$Rb is depicted in Fig. 1a. The common model for CPT with $\sigma^+$ light, describing CPT in the $m_F = 0$ resonance, reduces the atomic structure to a 3-level Λ-system. These levels represent the two ground 'clock states', ($F = 1$, $m_F = 0$) and ($F = 2$, $m_F = 0$), and one excited state, e.g., ($F' = 2$, $m_F = 1$). These are denoted as $|1\rangle$, $|2\rangle$, and $|3\rangle$, respectively, and are coupled by two $\sigma^+$-polarized electromagnetic fields. However, in order to allow for more elaborated processes, such as optical pumping, a more detailed model is required. Our model includes four additional states as illustrated in Fig. 1c. To account for the ground-state depopulation and for the $\sigma^+$ optical pumping, we represent the $m_F \neq 0$ states in the ground manifold by the effective state $|4\rangle$, denoted as the 'trap state'. To incorporate the re-pumping action of the repump-beams, we represent the $m_F \neq 0$ states in the excited levels by the effective state |5⟩, denoted as the 'π-excited state'. State |4⟩ is coupled to state |5⟩ by a π-polarized electromagnetic field. To account for off-resonance pumping of the repump-beams, as explained later, we represent the excited $m_F = 0$ states, ($F' = 1$, $m_F = 0$) and ($F' = 2$, $m_F = 0$), by two effective states, |6⟩ and |7⟩, denoted as the 'off-resonance pumping states'. Due to selection rules for the π-polarization, state |1⟩ is decoupled from state |6⟩, and state |2⟩ is decoupled from state |7⟩. Therefore, in our model, the π-polarized fields off-resonantly couple states |1⟩ to |7⟩ and |2⟩ to |6⟩. The excited states, |3⟩, |5⟩, |6⟩, and |7⟩, may spontaneously decay to all the lower levels in accordance to selection rules, and the ground states, |1⟩, |2⟩, and |4⟩, exhibit standard population transfer and decoherence.

The model Hamiltonian is composed of the atomic Hamiltonian, with a suitable energy level for each state, and a coupling Hamiltonian with two $\sigma^+$ and two π electromagnetic modes (examples for the allowed transitions are represented by arrows in Fig. 1c). Non-Hamiltonian decay is introduced using Lindblad terms within the Master equation formalism [17]. To describe depolarization or spontaneous emission and decoherence in the Lindblad form, we used the |i⟩⟨j| and |i⟩⟨i| operators, respectively. Finally, the steady-state solution of the system, in terms of a density operator, is obtained numerically, yielding the absorption coefficients for the various fields and the population distribution in the ground state.

The cell temperature and thus the atomic density were constant in all experiments and have been determined from absorption measurements. Some of the dipole matrix-elements and the decay rates in the model are effective parameters and thus required calibration. In practice, we have set them according to relevant known values of the 16-level system, and only used the Rabi frequency of the repump-beams as a fit parameter. Also, we have calibrated an effective frequency detuning of the off-resonant π-transitions from the relevant experiments.

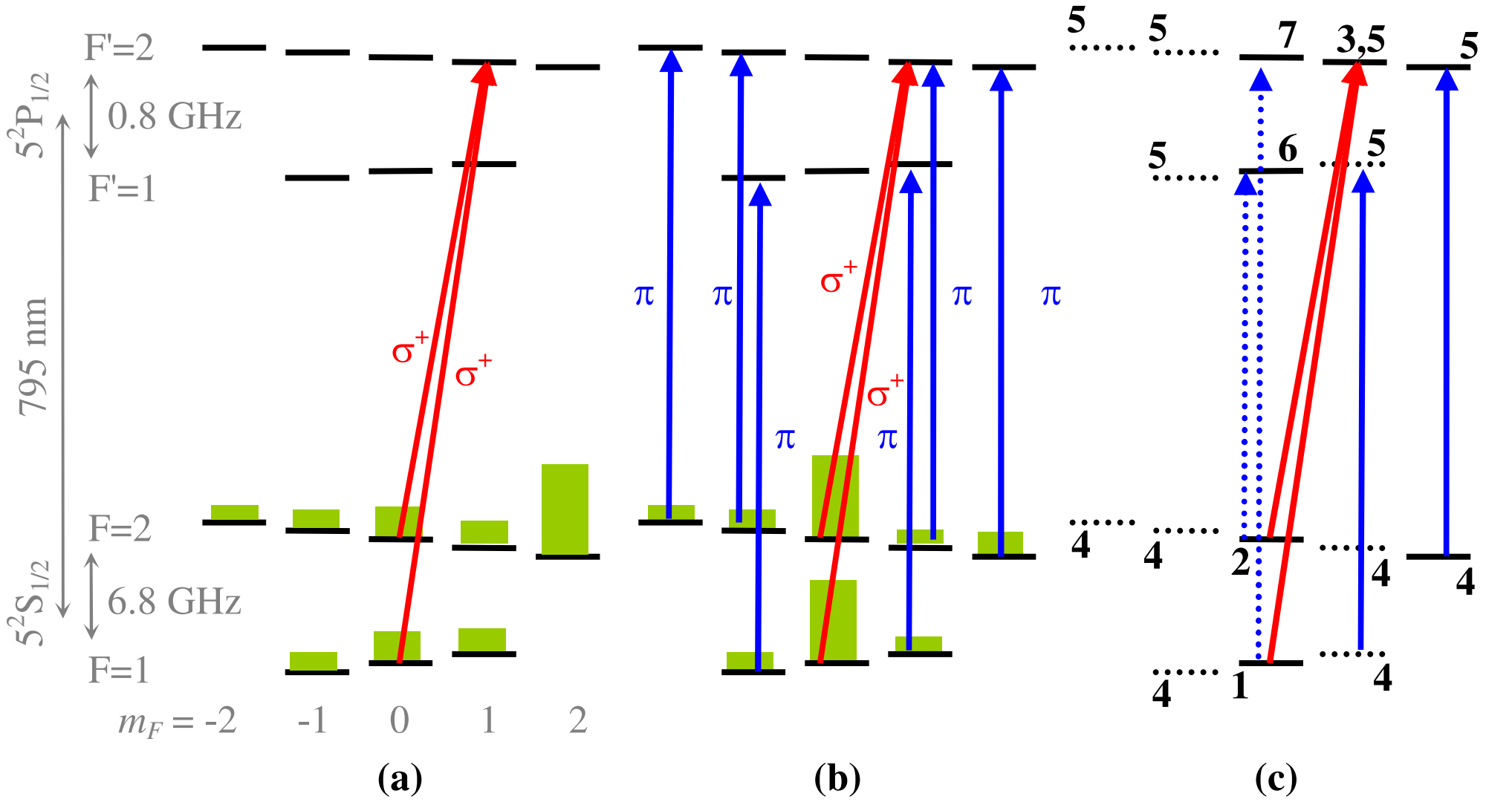

Fig. 1. (Color online) Energy levels diagram for CPT clock in $^{87}$Rb D1 line, without the repump-beams (a) and with the repump-beams (b). The green bars illustrate the population distribution. In (a), the $\sigma^+$-beams create the CPT resonance while pumping the population towards the end state ($m_F$=2). In (b), the π transitions ($F$ = 1, $m_F$ = 0) → ($F'$ = 1, $m_F$ = 0) and ($F$ = 2, $m_F$ = 0) → ($F'$ = 2, $m_F$ = 0) are forbidden, and thus the repump-beams re-pump the population to the $m_F$ = 0 states. In (c), the effective states and allowed transitions in the numerical model are marked: the CPT states (1, 2, and 3), the 'trap state' (4), the 'π-excited state' (5), and the 'off-resonance pumping states' (6 and 7). The dashed arrows represent the off-resonant π-excitations.

## III. The experimental setup

The experimental setup is depicted in Fig. 2 (left). A vertical-cavity surface-emitting laser diode (VCSEL) is stabilized to the *D1* transition of $^{87}$Rb (~795nm). The VCSEL is current-modulated at 3.417GHz, and the -1, +1 sidebands are tuned to the $F = 2 \rightarrow F' = 2$, $F = 1 \rightarrow F' = 2$ transitions, respectively (CPT-VCSEL in Fig. 2). The polarization of the laser is set to right-circular by a polarizing beam-splitter and a quarter-wave plate. The beam passes through a cross-shaped vapor cell along the *z*-direction, and the transmitted intensity is measured with a photo-diode. A second VCSEL (repump-VCSEL) is current modulated at ~3GHz, and its -1, +1 sidebands are tuned to the $F = 2 \rightarrow F' = 2$, $F = 1 \rightarrow F' = 1$ transitions, respectively. The repump-VCSEL is set to linear polarization parallel to the *z*-direction (π-polarization), and passes the vapor cell in perpendicular to the CPT beam. The intensity of both beams is controlled by neutral-density filters.
The cross-shaped vapor cell (Fig. 2, right) contains isotopically pure $^{87}Rb$ and 10 Torr of nitrogen buffer gas. The dimensions of the cell are 18mm length, along the *z*-direction, and 6mm diameter. Due to technical limitations, the overlap of the repump beams along optical path of the CPT beam is limited to 6mm. Therefore, only one third of the path within the cell is directly repumped (see Fig. 2). The temperature of the cell is controlled using electrical heaters (glass with ITO coating), setting the rubidium vapor density to $n$ = 7x10$^{10}$atoms/cc. The resulting small-signal transparency is about 40%, providing the optimal conditions for a CPT clock [18]. A four-layered magnetic shield is used to isolate the cell from the earth magnetic field. Three sets of Helmholtz coils are used to set a small magnetic field, $B_z$ = 20mG, along the z-direction, and to eliminate the residual magnetic field in the perpendicular plane.

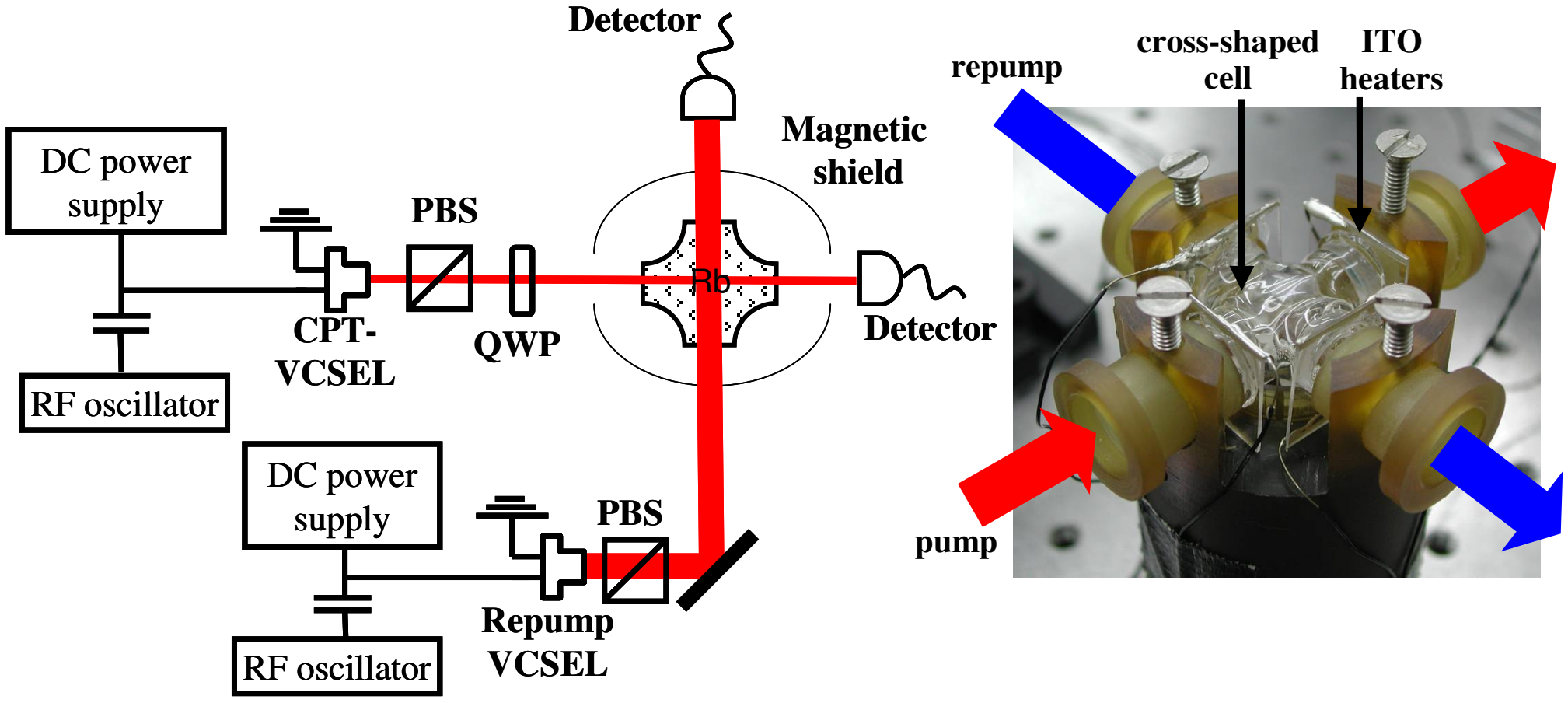

Fig. 2. Experimental setup. Left: VCSEL – vertical cavity surface emitting laser; PBS – polarizing beam-splitter; QWP – ¼-wave plate (Not shown: three sets of Helmholtz coils within the magnetic shield). Right: The cross-shaped glass vapor cell with Indium-Tin-Oxide (ITO) heaters pressed against its four facets.

## IV. Results

We first study the CPT spectrum of our experimental system while applying the repump beams. The spectrum is obtained by shutting-down the electrical heaters and scanning the frequency of the RF oscillator, which current modulates the CPT-VCSEL, around the CPT clock-transition. The scanning rate is sufficiently slow, so that the steady-state CPT spectrum is obtained. The repump-VCSEL is current modulated using a constant frequency of the RF oscillator. For five different CPT-VCSEL intensities, we scan the repump-VCSEL intensity and measure the CPT resonance. Each CPT line is fitted to a Voigt profile, and its full width at half maximum (FWHM) and contrast (the ratio between the CPT resonance amplitude and its background) are inferred. Figure 3 (left) depicts the measured CPT contrast of the spectrum versus the repump-VCSEL intensities for the five different CPT-VCSEL intensities.

When the intensity of the CPT-VCSEL is weak, the pumping to high $m_F$ states is small, and the repump effect is negligible. As the intensity of the CPT-VCSEL is increased, a larger fraction of the atoms is pumped to higher $m_F$ states, and the repump beams are more effective, bringing back more atoms to the $m_F = 0$ state and increasing the CPT contrast. The effectiveness of the repump beams depends also on their intensity. Obviously at low intensities, the repump effect is small, and it increases with the intensity. However at larger intensities, the improvement in the contrast diminishes, and eventually the contrast begins to drop. We attribute this to the off-resonant pumping from the $m_F$=0 states. The solid lines in Fig. 3 are the results of our numerical model, showing a good agreement with the experimental results. Figure 3 (right) depicts the measured CPT linewidth (FWHM) versus the repump-VCSEL intensity at five different CPT-VCSEL intensities. The CPT lines broaden with the laser intensity due to power broadening.

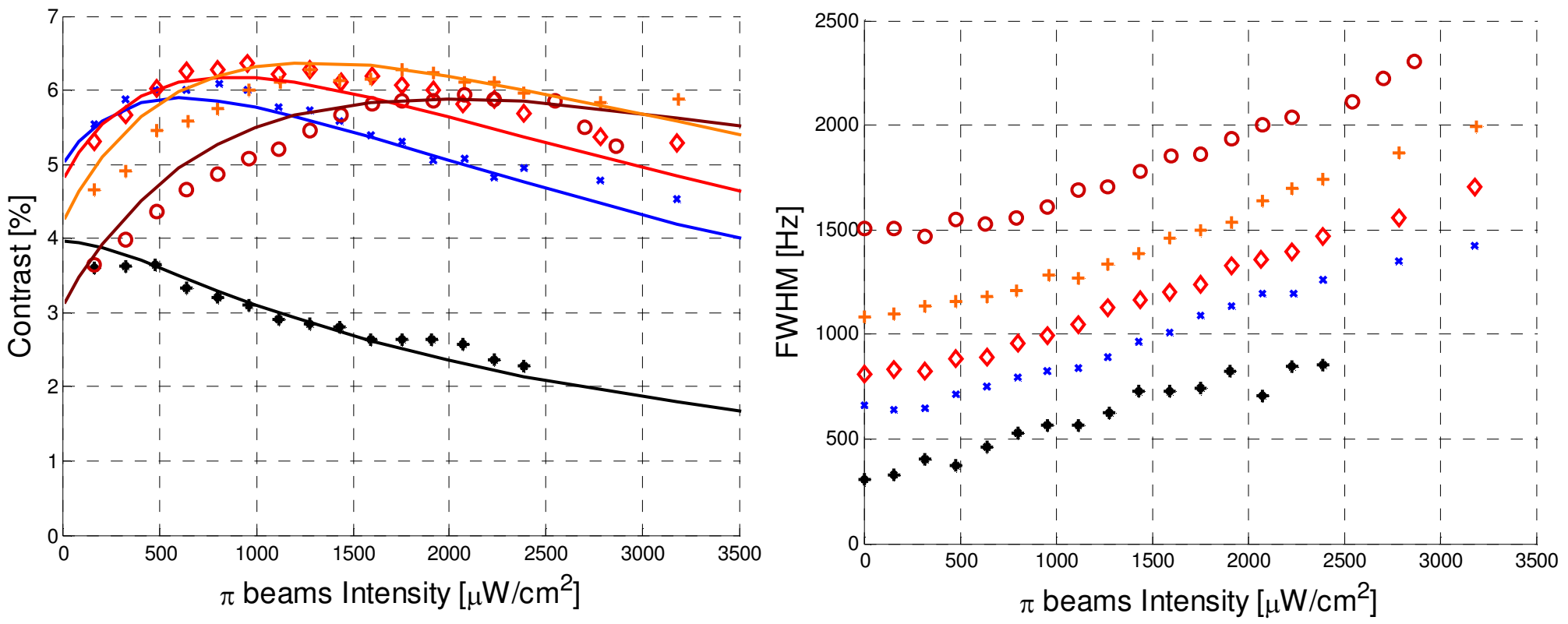

Fig.3. (Color online) The CPT contrast (left) and FWHM (right) as a function of the repump-VCSEL intensity at different CPT-VCSEL intensities (* Black 1440μW/cm2, x Blue 4320μW/cm2, ◊ Red 5760μW/cm2, + Orange 8640μW/cm2, ○ Brown 12960μW/cm2). The lines are our model's results.

A second set of measurement was devoted to verify the effectiveness of the repump beams, by studying the CPT spectrum *without* applying the repump beams. Figure 4 depicts the measured properties of the CPT spectrum at various laser intensities. The CPT linewidth increases linearly with the laser intensity due to power broadening. The CPT contrast initially increases with the laser intensity, but then decreases as the population is pumped out of the clock transition into the high $m_F$ states [3], in good agreement with our numerical model. This phenomenon poses a fundamental limit on the maximal contrast that can be obtained in this CPT apparatus and limits the possible accuracy of CPT based frequency standards.

The maximal increase in contrast achieved using the repump beams, for different CPT-VCSEL intensities, is also plotted in Fig. 4 (along with the associated FWHM). It is important to note that, due to technical limitations in the current apparatus, the repump beam is applied to only one third of the length of the cell, limiting the effectiveness of the repumping method.

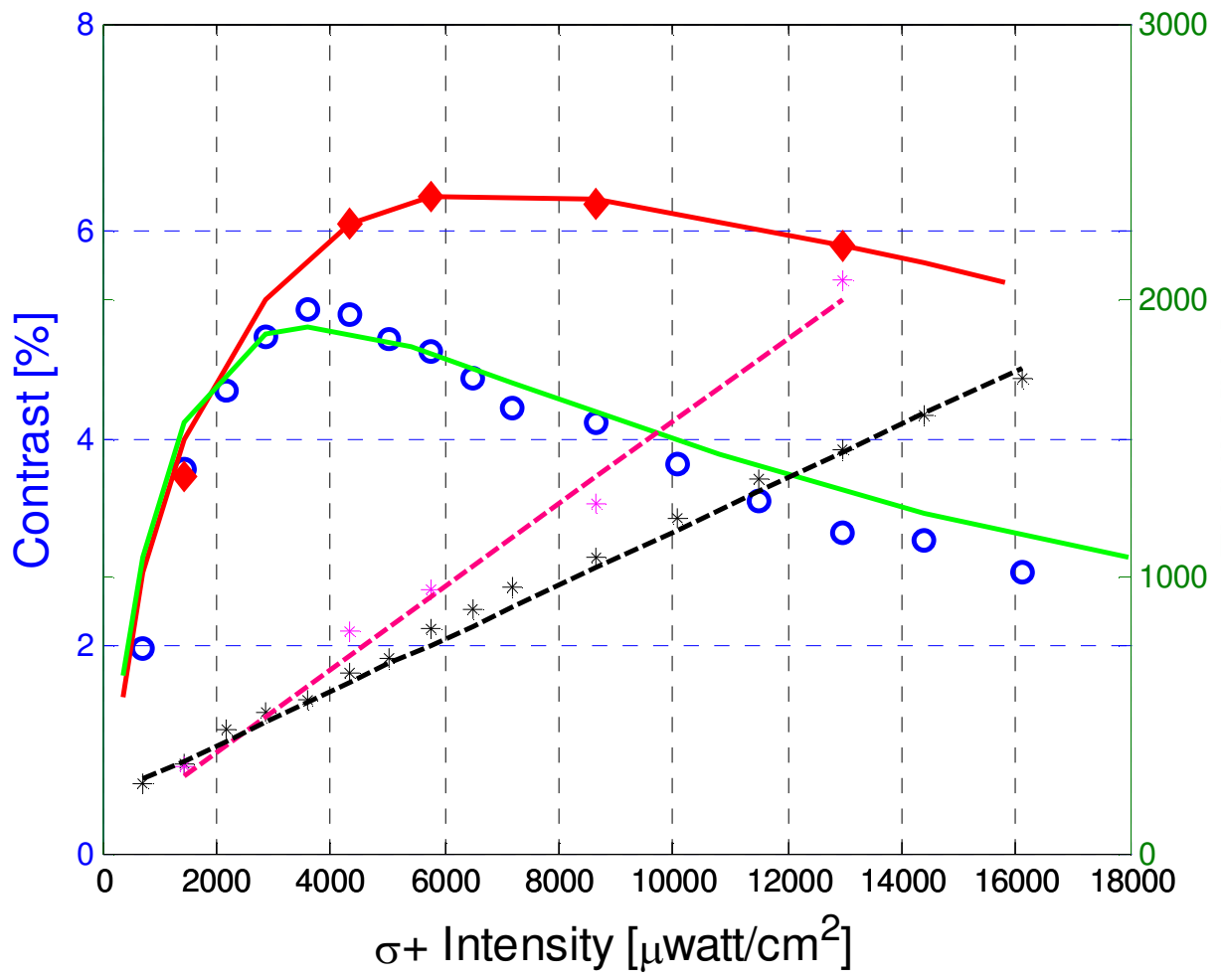

Fig.4. (Color online) The CPT contrast (blue circles) and the CPT FWHM (black asterisks) without the repump-beams, as a function of the CPT beams intensity. The red diamonds and the magenta asterisks are the maximum contrasts and the FWHM, obtained with the repump-beams, respectively. The green and the red lines are the model results.

To obtain the best frequency stability, it is important to maximize the contrast and to minimize the linewidth [1]. Therefore, in order to use this optical pumping technique to substantially improve the short-term frequency stability, it is necessary to pump the atoms without increasing the CPT linewidth due to power broadening. One option is to time-alternate the repump beams, and perform the CPT measurement when the repump beams are off. Another possibility is to apply the repump beams near the vicinity of the CPT beams, but with minimal spatial overlap. In that case the repumping effect will be obtained utilizing the diffusion of atoms in the cell [19].

The contrast observed at the optimum operating point of the system is ~6.1% using the repump beams, instead of 5% without repump beams. For an arrangement of maximal spatial overlap of the two beams, our numerical model predicts a contrast of about 20%, demonstrating an improvement of a factor of 4.

## V. Conclusion

We have demonstrated an optical pumping technique to increase the contrast of CPT-based frequency standards utilizing two additional optical modes (repump beams). The improvement of the CPT contrast was studied for a wide range of CPT and repump beams intensities. A major drawback of this pumping method is the power broadening of the CPT resonance associated with the repump beams. This limitation may be prevented by turning off the repump beam just before making the measurements or by applying them in the near vicinity of the CPT beams, with minimal spatial overlap. We estimate that by using an optimal setup, the contrast may be improved by up to a factor of four.